\documentclass{PoS}
\usepackage{bm}
\newcommand{\pvec}{\bm{p}}
\newcommand{\xvec}{\bm{x}}
\newcommand{\yvec}{\bm{y}}
\newcommand{\rvec}{\bm{r}}

\title{Exploring Excited Hadrons}

\ShortTitle{Exploring Excited Hadrons}

\author{\speaker{Colin Morningstar}\\ 
        Department of Physics, Carnegie Mellon University, 
        Pittsburgh, PA 15213, USA\\
        E-mail: \email{colin\_morningstar@cmu.edu}}

\abstract{Progress in extracting the spectrum of excited hadron resonances is
reviewed and the key issues and challenges in such computations are outlined.
The importance of multi-hadron states as simulations are done with lighter pion
masses is discussed, and the need for all-to-all quark propagators is 
emphasized.}

\FullConference{The XXVI International Symposium on Lattice Field Theory \\
		 July 14 - 19, 2008\\
		 Williamsburg, Virginia, USA}

\begin{document}

\section{Introduction}
Experiments show that many excited-state hadrons exist, and there are significant
experimental efforts to map out the QCD resonance spectrum, such as Hall B
and the proposed Hall D at Jefferson Lab, ELSA associated with the University
of Bonn, COMPASS at CERN, PANDA at GSI, and BESIII in Beijing.
Hence, there is a great need for \textit{ab initio} determinations of such
states in lattice QCD.

Higher-lying excited hadrons are a new frontier in lattice QCD, and
explorations of new frontiers are usually fraught with dangers.  Excited states 
are more difficult to extract in Monte Carlo calculations; correlation matrices 
are needed and operators with very good overlaps onto the states of interest are
crucial.  To study a particular state of interest, all states lying below that
state must first be extracted, and as the pion gets lighter in lattice QCD 
simulations, more and more multi-hadron states will lie below the excited
resonances.  To reliably extract these multi-hadron states, multi-hadron 
operators made from constituent hadron operators with well-defined relative
momenta will most likely be needed, and the computation of temporal correlation
functions involving such operators will require the use of all-to-all quark 
propagators.  The evaluation of disconnected diagrams will ultimately be
required.  Perhaps most worrisome, most excited hadrons are unstable 
(resonances), so the results obtained for finite-box stationary-state
energies must be interpreted carefully.

This talk will describe the key issues and challenges in exploring excited
hadrons in lattice QCD, emphasizing the importance of multi-hadron operators
and the need for all-to-all quark propagators.  Dealing with resonances in a
box is discussed, and the technology associated with extracting excited
stationary-state energies, including operator design and field smearing,
is detailed.  Efforts in variance reduction of stochastically-estimated
all-to-all quark propagators using source dilutions are outlined.  Results
on excited hadrons during the last year are summarized.

\section{Resonances in a box}
\begin{figure}[t]
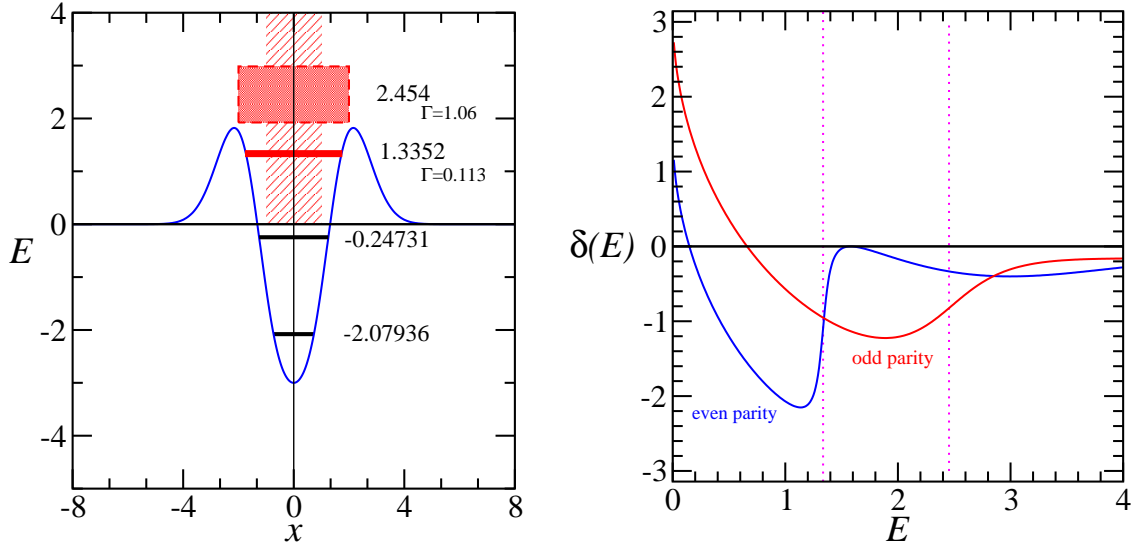

\begin{center}
\includegraphics[width=2.7in,bb=64 43 528 531]{spectrum_inf.eps}
\hspace*{4mm}
\includegraphics[width=2.95in,bb=15 36 528 523]{delta_inf.eps}
\end{center}
\caption[one]{(Left) The solid curve shows the potential $V(x)$ of the
example given in Eq.~(\ref{eq:example}).  The spectrum for energies $E<4$ is also 
shown.  The ground state is an even-parity bound state, and there is one bound 
state in the odd-parity channel.  Two resonances (a narrow one in the even-parity
and a very broad one in the odd parity sector) in a continuum of
scattering states $E>0$ are shown. (Right) The scattering phases 
$\delta_\pm(E)$ for this example are shown for $E<4$.  The locations of the
resonance energies are shown as dashed vertical lines.
\label{fig:one}}
\end{figure}

A simple example serves to illustrate the issues that must be confronted
when studying resonances in a box.  Consider the (dimensionless) Hamiltonian 
for a single nonrelativistic particle of mass $m=1$ moving in one dimension 
in a potential $V(x)$ is given by
\begin{equation}
   H = \textstyle\frac{1}{2}p^2 + V(x),\qquad
  V(x)= (x^4-3)\ e^{-x^2/2}.
\label{eq:example}
\end{equation}
This potential has an attractive core surrounded by a repulsive barrier,
as shown in Fig.~\ref{fig:one}.  The infinite-volume spectrum of this
Hamiltonian for energies $E<4$ is shown in Fig.~\ref{fig:one}.  The
ground state is a bound state of even parity, and there is one bound
state in the odd parity channel.  A continuum of scattering states is
found for $E>0$, with a narrow resonance in the even-parity channel and
a broad resonance in the odd-parity sector, both below $4$.  Even $(+)$ and
odd $(-)$ parity scattering phase shifts $\delta_\pm(E)$ can be defined 
in the usual way as the phase between the transmitted and incident wave,
appearing in the asymptotic wave functions as
\begin{equation}
    \varphi_k^{(+)}(x) = c_+\cos\biggl(k\vert x\vert + \delta_+(k)\biggr),\quad
    \varphi_k^{(-)}(x) = c_-{\rm sgn}(x)\sin\biggl(k\vert x\vert 
  + \delta_-(k)\biggr),\quad
 (\vert x\vert\rightarrow\infty),
\end{equation}
where $k=\sqrt{2E}$.
These phase shifts are also shown in Fig.~\ref{fig:one}.  The narrow
even-parity resonance is seen as a sudden dramatic increase in the phase 
shift (by an amount comparable to $\pi$) and the broad odd-parity 
resonance appears as a not-so-sudden increase in the phase shift.  The
resonance masses and widths can be extracted by fitting a Breit-Wigner
plus a polynomial background to $d\delta_\pm/dE$ in the vicinity of the
resonances (or by employing the complex rotation method described in
Ref.~\cite{crm1}).

Now consider solving the system in a box of length $L$ such that
$-\frac{1}{2}L<x<\frac{1}{2}L$, assuming periodic boundary 
conditions.  Note that the potential is now
$
  V_L(x) = \sum_{n=-\infty}^\infty V(x+nL).
$
The infinite volume gets tiled into $L$-length strips in which
the potential is replicated. We assume that $L$ is large compared to the
extent of the potential $V$ so that interactions with mirror
potentials is negligible.

\begin{figure}
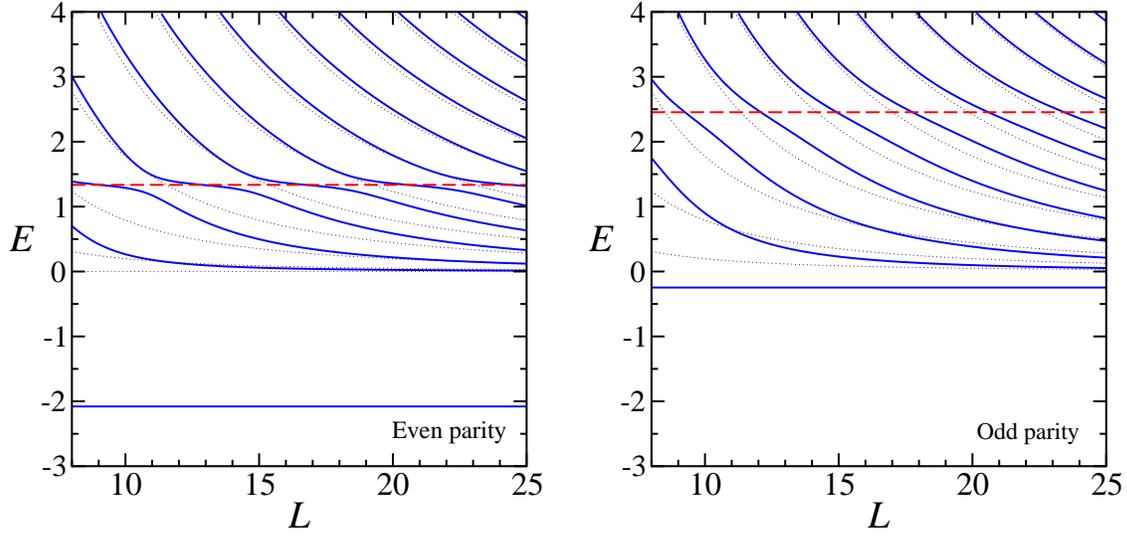

\begin{center}
\includegraphics[width=2.8in,bb=32 73 536 531]{spectrum_even_lat.eps}
\hspace*{4mm}
\includegraphics[width=2.8in,bb=32 73 536 531]{spectrum_odd_lat.eps}
\end{center}
\caption[two]{
(Left) Spectrum of even-parity states for the example Hamiltonian
given in Eq.~(\ref{eq:example}) in a box of extent $L$ with periodic
boundary conditions.  (Right) Spectrum of odd-parity states.
The light dotted lines indicate the energy levels for $V=0$ which have
values $2\pi^2 n^2/L^2$ for $n=1,2,3,\dots,$ plus an $n=0$ line in the
even parity channel.  The horizontal dashed lines show the locations
of the infinite-volume resonance energies.
\label{fig:two}}
\end{figure}

In a finite-box with periodic boundary conditions, the momentum
is quantized, so the entire spectrum is a series of discrete energies,
even for $E>0$.  The periodic-box spectrum can be determined in two
ways: diagonalization of the Hamiltonian in a basis of states
having appropriate boundary conditions, and by solving the differential
equation and matching to an asymptotic form having the correct
boundary conditions.  Either way, one finds the spectrum shown in
Fig.~\ref{fig:two}.   The light dotted lines indicate the spectrum
for $V=0$ which have values $2\pi^2 n^2/L^2$ for $n=1,2,3,\dots,$ plus
an $n=0$ line in the even parity channel.  One sees that a resonance
shows up as a series of avoided level crossings when viewed against
box length $L$.  A narrow
resonance, as in the even parity sector, can be easily identified by a
closely-avoided level crossing, but
a broad resonance, as in the odd parity sector, is essentially impossible
to recognize.

\begin{figure}[b]
\begin{center}
\includegraphics[width=2.7in,bb=14 54 338 301]{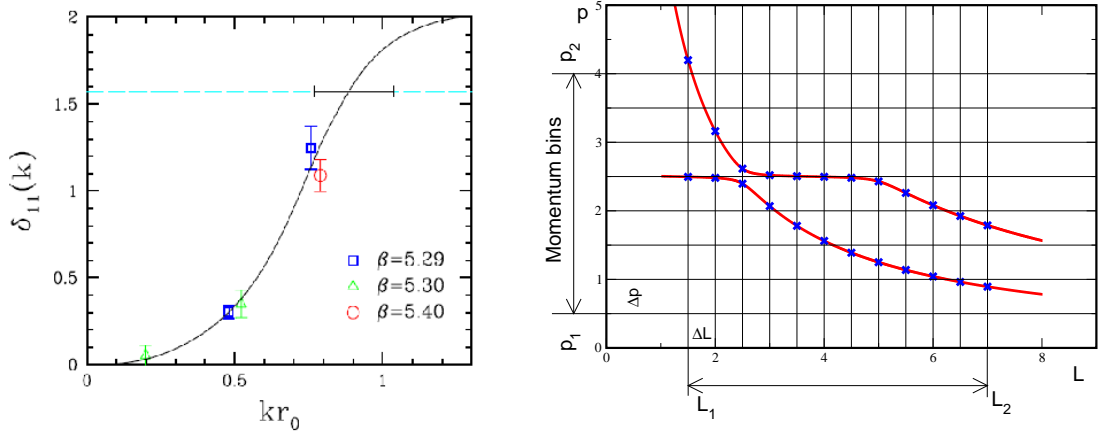}
\hspace*{4mm}
\includegraphics[width=2.9in,bb=15 50 707 528]{binning.eps}
\end{center}
\caption[three]{
(Left) $\pi\pi$ scattering phase shift showing the $\rho$-resonance
\cite{schierholz}. A width $\Gamma_\rho=200^{+130}_{-100}$~MeV is
obtained from a Breit-Wigner fit. (Right) Illustration of the binning method
of Ref.~\cite{bernard} in which a probability distribution is used
to identify resonant structure.  For various values of $L$, energies
are collected into momentum bins to construct a probability 
distribution $W(p)$.
\label{fig:three}}
\end{figure}

These plots illustrate the difficulty in extracting resonance parameters
from finite-box energies.  Under certain special circumstances, resonance
parameters can be ferreted out using tricks such as described in 
Ref.~\cite{mcneile}.  Examining the spectrum in several volumes is
important, and knowing the pattern of multi-hadron states based on
mass determinations of the stable particles and group-theoretical
combinations of the constituents having total zero-momentum certainly 
helps.  For high precision, the phase-shift method of 
Refs.~\cite{dewitt,luscher} can be used.  In this method, the finite-volume
energies are used to determine the scattering phase shifts of the partial
waves, from which one can
deduce resonance masses and widths.  This method has recently been
applied to study the $\rho$-meson resonance\cite{schierholz}.  The
$\pi\pi$ phase shift was extracted from the finite-volume spectrum,
and the $\rho$ resonance parameters extracted (see Fig.~\ref{fig:three}).
However, the practicality of the phase-shift method has yet to be
demonstrated for other resonances in QCD.

\begin{figure}
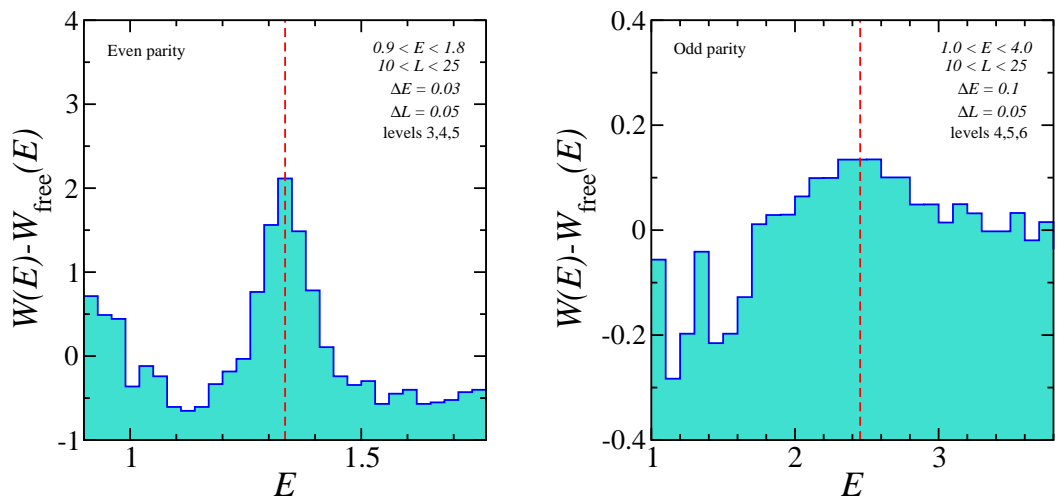

\begin{center}
\includegraphics[width=2.5in,bb=36 33 523 531]{even_binning_subt.eps}
\hspace*{10mm}
\includegraphics[width=2.5in,bb=36 33 523 532]{odd_binning_subt.eps}
\end{center}
\caption[four]{
(Left) Results of applying the binning method of Ref.~\cite{bernard}
near the even-parity resonance of the example Hamiltonian in 
Eq.~(\ref{eq:example}).  Location of the resonance is indicated
by the vertical dashed line. (Right) Results for the very broad
odd-parity resonance. 
\label{fig:four}}
\end{figure}

A new histogram method has been recently proposed\cite{bernard} and applied
to synthetic pion-nucleon data which mocks up the $\Delta$-resonance.
Energies for several volumes are collected into momentum bins, and with
suitable normalization and free-result subtraction, such histograms produce
a probability distribution $W(p)$ which shows peaks corresponding to the
resonances of interest (see Fig.~\ref{fig:three}).  The method has no
prior theoretical bias and provides the possibility of seeing resonant
structure even when avoided level crossings are washed out by a broad
resonance.  The application of this method to the example Hamiltonian
in Eq.~(\ref{eq:example}) is shown in Fig.~\ref{fig:four}.  The narrow
resonance appears quite clearly, and amazingly, the odd-parity broad 
resonance is also correctly reproduced.

Deducing resonance parameters from finite-box spectra remains a difficult
challenge, especially considering that higher-lying resonances will
lie above three-particle and four-particle thresholds and that these
resonances can have multiple decay channels.  Certainly, further work
in this area is needed.  Perhaps matching the finite-box spectra to 
that of an effective theory, such as a one-boson exchange model, might
ultimately be the way to make progress.

\section{Excited stationary states: recent results}

Before discussing the issues and challenges in extracting the
energies of stationary states in a box, I would like to summarize
the excited-state results which have appeared since the last lattice
conference.

\begin{figure}
\begin{center}
\includegraphics[width=2.68in,bb=6 44 532 662]{nucleon_spectrum_lat.eps}
\hspace*{10mm}
\includegraphics[width=2.5in,bb=156 220 421 555]{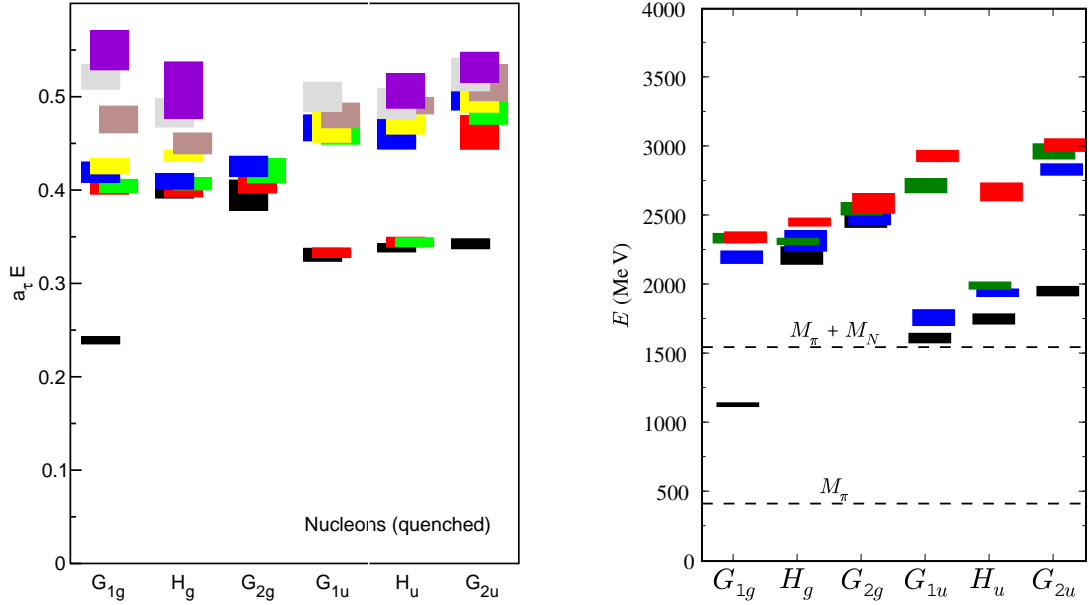}
\end{center}
\caption[nucleons]{
(Left) Nucleon spectrum from 200 quenched configurations
on a $12^3\times 48$ anisotropic lattice using the Wilson gauge and quark
actions with $a_s \sim 0.1$ fm,
$a_s/a_t \sim 3.0$ and $m_\pi\sim 700$~MeV from Ref.~\cite{adamthesis}.  
(Right) Nucleon spectrum from 430 $N_f=2$ configurations on a $24^3\times 64$
lattice using a stout-smeared clover fermion action and Symanzik-improved
gauge action with $a_s\sim 0.1$~fm, $a_s/a_t\sim 3$, and $m_\pi=400$~MeV
from Ref.~\cite{nucleons2}.
\label{fig:nucleonspectra}}
\end{figure}

A first glimpse of the higher-lying nucleon spectrum in lattice QCD was
provided by the Hadron Spectrum Collaboration in Ref.~\cite{adamthesis}.
These first results, shown in Fig.~\ref{fig:nucleonspectra}, were on a small
$12^3\times 48$ anisotropic quenched lattice with a very heavy pion.
Results for both the nucleons and $\Delta$-resonances on
239 quenched configurations on a $16^3\times 64$ lattice and 167 quenched
configurations on a $24^3\times 64$ lattice using an anisotropic Wilson
action with spatial spacing $a_s\sim 0.1$~fm, $a_s/a_t\sim 3$, and a
pion mass $m_\pi\sim 490$~MeV appeared during the past 
year\cite{nucleons1}.  These masses have been determined in the
past year using 430 $N_f=2$ configurations on a $24^3\times 64$ 
lattice with a stout-smeared 
clover fermion action and a Symanzik-improved anisotropic gauge 
action\cite{nucleons2}.  The results for a pion mass $m_\pi=400$~MeV,
spacing $a_s\sim 0.1$~fm and $a_s/a_t\sim 3$
are shown in Fig.~\ref{fig:nucleonspectra}.  The low-lying odd-parity
band shows the exact number of states in each channel as expected
from experiment.  The two figures show the splittings in the band
increasing as the quark mass is decreased.  At these heavy pion masses,
the first excited state in the $G_{1g}$ channel is significantly higher
than the experimentally measured Roper resonance.  It remains to be
seen whether or not this level will drop down as the pion mass is further
decreased.  Most of the levels in the right-hand plot lie very close
to two-particle thresholds.  The use of two-hadron operators will be
needed to go to lighter pion masses.

\begin{figure}
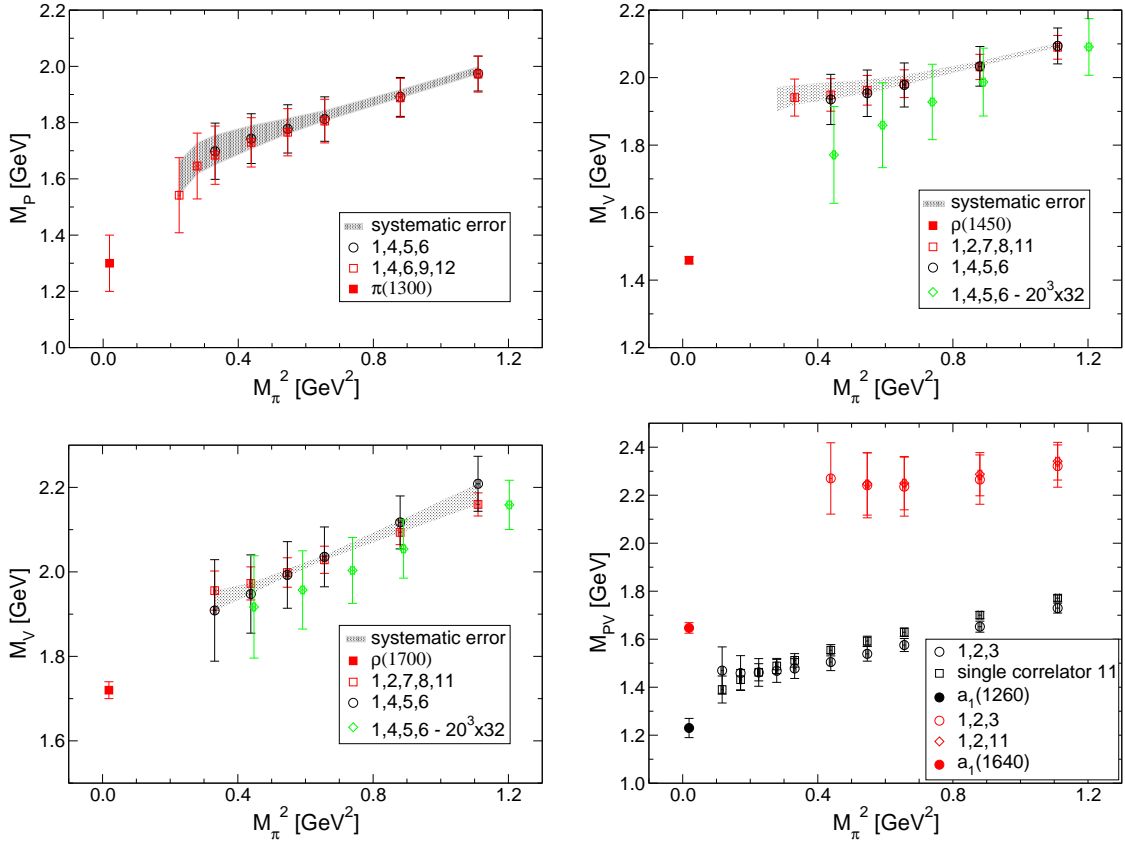

\begin{center}
\includegraphics[width=2.8in,bb=9 14 736 561]{gatt1.eps}
\hspace*{4mm}
\includegraphics[width=2.8in,bb=9 14 736 561]{gatt2.eps}\\
\includegraphics[width=2.8in,bb=10 14 736 553]{gatt3.eps}
\hspace*{4mm}
\includegraphics[width=2.8in,bb=9 14 736 603]{gatt4.eps}
\end{center}
\caption[gatt]{
(Upper left) Results from Ref.~\cite{gatt} for the first-excited 
pseudoscalar $0^{-+}$ meson mass against pion mass squared from 99 quenched 
configurations on a $16^3\times 32$ lattice with a chirally-improved 
fermion action and the Luscher-Weisz gauge action for lattice spacing 
$a_s\sim 0.15$~fm. 
(Upper right) First-excited $1^{--}$ meson mass. (Lower left) Second-excited
$1^{--}$ meson mass. (Lower right) Ground and first-excited $1^{++}$
meson masses.
\label{fig:gatt}}
\end{figure}

During the past year, extractions of excited meson states have been
presented in Ref.~\cite{gatt}.  Results in the pseudoscalar, vector,
and axial-vector channels are shown in Fig.~\ref{fig:gatt}.  These
results were obtained using 99 quenched configurations on a $16^3\times 32$
isotropic lattice with a chirally-improved fermion action and the
Luscher-Weisz gauge action for lattice spacing $a_s\sim 0.15$~fm
and a range of pion masses.  This work emphasizes the use of
derivative sources in correlation matrices to obtain the excited
states.  A search for light scalar tetraquark states with isospin
$I=0,\frac{1}{2}$ was also presented at this conference\cite{tetraquark}.

\section{Excited stationary states: key issues}

Reliably capturing the masses of excited states requires the computation
of correlation matrices $C_{ij}(t)=\langle 0\vert T \Phi_i(t)
 \Phi^\dagger_j(0) \vert 0\rangle$ associated with a large set of $N$
different operators $\Phi_i(t)$.  It has been shown in Ref.~\cite{wolff90}
that the $N$ {\em principal effective masses} $W_\alpha(t)$, defined by
\[  W_\alpha(t)=\ln\left(\frac{\lambda_\alpha(t,t_0)}{
 \lambda_\alpha(t+1,t_0)}\right),
\]
where $\lambda_\alpha(t,t_0)$ are the eigenvalues of
 $C(t_0)^{-1/2}\ C(t)\ C(t_0)^{-1/2}$ and $t_0<t/2$ is usually chosen,
tend to the eigenenergies of the lowest $N$ states with which the
$N$ operators overlap as $t$ becomes large.  The eigenvectors
associated with $\lambda_\alpha(t,t_0)$ can be viewed as 
variationally optimized operators.  When combined
with appropriate fitting and analysis methods, such variational
techniques are a particularly powerful tool for investigating excitation
spectra.  To extract the stationary state energies, one can fit
a single-exponential or a sum of two exponentials to each principal
correlator; alternatively, optimized operators can be determined
on an early time slice, and a fit to the correlation matrix of 
optimized operators can be carried out.  The use of both methods
is a good consistency check.

The use of operators whose correlation functions $C(t)$ attain their
asymptotic form as quickly as possible is crucial for reliably
extracting excited hadron masses.  An important ingredient in constructing
such hadron operators is the use of smeared fields.  Operators constructed
from smeared fields have dramatically reduced mixings with the high frequency
modes of the theory.  Both link-smearing and quark-field smearing should
be applied.  Since excited hadrons are expected to be large objects, 
the use of spatially extended operators is another key ingredient in
the operator design and implementation.  A more detailed discussion
of these issues can be found in Ref.~\cite{baryons1}.

Spatial links can be smeared using the stout-link procedure described in
Ref.~\cite{stout}. The stout-link smearing scheme is analytic, efficient,
and produces smeared links which are automatically elements of $SU(3)$ 
without the need for a projection back into $SU(3)$.  
Note that only spatial staples are used in the link smoothening; no temporal
staples are used, and the temporal link variables are not smeared.
The smeared quark fields can be defined by
\begin{equation}
\widetilde{\psi}(x) = \left(1+\frac{\sigma_s^2}{4n_\sigma}\ \widetilde{\Delta}
 \right)^{n_\sigma}\psi(x),
\end{equation}
where $\sigma_s$ and $n_\sigma$ are tunable parameters ($n_\sigma$ is 
a positive integer) and the three-dimensional covariant Laplacian
operators are defined in terms of the smeared link variables 
$\widetilde{U}_j(x)$ as follows: 
\begin{eqnarray}
 \widetilde{\Delta} O(x) &=& \!\!\!\sum_{k=\pm 1,\pm2,\pm3} \biggl(
  \widetilde{U}_k(x)O(x\!+\!\hat{k})-O(x) \biggr), 
\end{eqnarray}
where $O(x)$ is an operator defined at lattice site
$x$ with appropriate color structure, and noting that
$\widetilde{U}_{-k}(x)=\widetilde{U}_k^\dagger(x\!-\!\hat{k})$.  
The smeared fields $\widetilde{\psi}$ and $\widetilde{\overline{\psi}}$ are 
Grassmann-valued; in particular, these fields anticommute in the same way
that the original fields do, and the square of each smeared field vanishes.

Hadron states are identified by their momentum $\pvec$, intrinsic
spin $J$, projection $\lambda$ of this spin onto some axis,
parity $P=\pm 1$, and quark flavor content (isospin, strangeness, {\it etc.}).
Some mesons also include $G$-parity as an identifying quantum number.
If one is interested only in the masses of these states, one can restrict
attention to the $\pvec=\bm{0}$ sector, so operators must be invariant 
under all spatial translations allowed on a cubic lattice.  The little 
group of all symmetry transformations on a cubic lattice which leave 
$\pvec=\bm{0}$ invariant is the octahedral point group $O_h$, so operators
may be classified using the irreducible representations (irreps) of $O_h$.
For mesons, there are ten irreducible representations $A_{1g}, A_{2g}, 
E_g, T_{1g}, T_{2g}, A_{1u}, A_{2u}, E_u, T_{1u}, T_{2u}.$  The 
representations with a subscript $g (u)$ are even (odd) under parity.
The $A$ irreps are one dimensional, the $E$ irreps are two dimensional,
and the $T$ irreps are three-dimensional.
The $A_1$ irreps contain the $J=0,4,6,8,\dots$ states, the $A_2$ irreps
contain the $J=3,6,7,9,\dots$ states, the $E$ irreps contain the
$J=2,4,5,6,7,\dots$ states, the $T_1$ irreps contain the
spin $J=1,3,4,5,\dots$ mesons, and the $T_2$ irreps contain the spin 
$J=2,3,4,5,\dots$ states.  For baryons, there are four two-dimensional irreps 
$G_{1g}, G_{1u}, G_{2g}$, $G_{2u}$ and two four-dimensional representations 
$H_g$ and $H_u$.  The $G_1$ irrep contains the
$J=\frac{1}{2},\frac{7}{2},\frac{9}{2},\frac{11}{2},\dots$ states,
the $H$ irrep contains the $J=\frac{3}{2},\frac{5}{2},\frac{7}{2},
\frac{9}{2},\dots$ states, and the $G_2$ irrep contains the 
$J=\frac{5}{2},\frac{7}{2},\frac{11}{2},\dots$ states.
The continuum-limit spins $J$ of our states must be deduced by examining
degeneracy patterns across the different $O_h$ irreps.

The authors of Ref.~\cite{baryons1} advocate operators designed
with one eye towards maximizing overlaps with the low-lying states of
interest, and the other eye towards minimizing the number of sources
needed to calculate the required quark propagators.  They emphasize that
a construction method which can be easily adapted for baryons, mesons, 
hybrid states, and multi-hadron systems is ideal.  Since the 
calculation of quark propagators can be computationally expensive, 
baryon, meson, and multiquark operators which share the same basic
building blocks is recommended.  

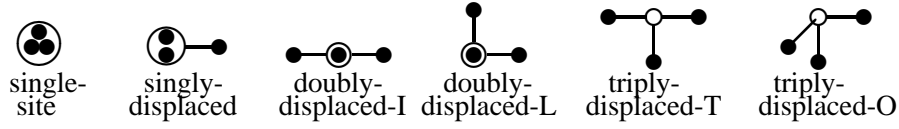
\begin{figure}[t]
\centerline{
\raisebox{0mm}{\setlength{\unitlength}{1mm}
\thicklines
\begin{picture}(16,10)
\put(8,6.5){\circle{6}}
\put(7,6){\circle*{2}}
\put(9,6){\circle*{2}}
\put(8,8){\circle*{2}}
\put(4,0){single-}
\put(5,-3){site}
\end{picture}}
\raisebox{0mm}{\setlength{\unitlength}{1mm}
\thicklines
\begin{picture}(16,10)
\put(7,6.2){\circle{5}}
\put(7,5){\circle*{2}}
\put(7,7.3){\circle*{2}}
\put(14,6){\circle*{2}}
\put(9.5,6){\line(1,0){4}}
\put(4,0){singly-}
\put(2,-3){displaced}
\end{picture}}
\raisebox{0mm}{\setlength{\unitlength}{1mm}
\thicklines
\begin{picture}(20,8)
\put(12,5){\circle{3}}
\put(12,5){\circle*{2}}
\put(6,5){\circle*{2}}
\put(18,5){\circle*{2}}
\put(6,5){\line(1,0){4.2}}
\put(18,5){\line(-1,0){4.2}}
\put(6,0){doubly-}
\put(4,-3){displaced-I}
\end{picture}}
\raisebox{0mm}{\setlength{\unitlength}{1mm}
\thicklines
\begin{picture}(20,13)
\put(8,5){\circle{3}}
\put(8,5){\circle*{2}}
\put(8,11){\circle*{2}}
\put(14,5){\circle*{2}}
\put(14,5){\line(-1,0){4.2}}
\put(8,11){\line(0,-1){4.2}}
\put(4,0){doubly-}
\put(1,-3){displaced-L}
\end{picture}}
\raisebox{0mm}{\setlength{\unitlength}{1mm}
\thicklines
\begin{picture}(20,12)
\put(10,10){\circle{2}}
\put(4,10){\circle*{2}}
\put(16,10){\circle*{2}}
\put(10,4){\circle*{2}}
\put(4,10){\line(1,0){5}}
\put(16,10){\line(-1,0){5}}
\put(10,4){\line(0,1){5}}
\put(4,0){triply-}
\put(1,-3){displaced-T}
\end{picture}}
\raisebox{0mm}{\setlength{\unitlength}{1mm}
\thicklines
\begin{picture}(20,12)
\put(10,10){\circle{2}}
\put(6,6){\circle*{2}}
\put(16,10){\circle*{2}}
\put(10,4){\circle*{2}}
\put(6,6){\line(1,1){3.6}}
\put(16,10){\line(-1,0){5}}
\put(10,4){\line(0,1){5}}
\put(4,0){triply-}
\put(2,-3){displaced-O}
\end{picture}}  }
\vspace*{8pt}
\caption{The spatial arrangements of the extended three-quark baryon
operators. Smeared quark-fields are
shown by solid circles, line segments indicate
gauge-covariant displacements, and each hollow circle indicates the location
of a Levi-Civita color coupling.  For simplicity, all displacements
have the same length in an operator.  Results presented here used
displacement lengths of $3a_s$ ($\sim 0.3 \mbox{ fm}$).
\label{fig:operators}}
\end{figure}

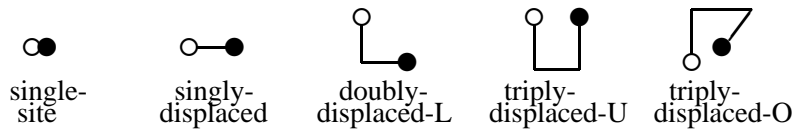
\begin{figure}[b]
\centerline{
\raisebox{0mm}{\setlength{\unitlength}{1mm}
\thicklines
\begin{picture}(20,12)
\put(7,7){\circle{2}}
\put(9,7){\circle*{2.5}}
\put(4,0){single-}
\put(5,-3){site}
\end{picture}}
\raisebox{0mm}{\setlength{\unitlength}{1mm}
\thicklines
\begin{picture}(20,12)
\put(6,7){\circle{2}}
\put(12,7){\circle*{2.5}}
\put(7,7){\line(1,0){4}}
\put(4,0){singly-}
\put(2,-3){displaced}
\end{picture}} 
\raisebox{0mm}{\setlength{\unitlength}{1mm}
\thicklines
\begin{picture}(20,12)
\put(7,11){\circle{2}}
\put(13,5){\circle*{2.5}}
\put(12,5){\line(-1,0){5}}
\put(7,10){\line(0,-1){5}}
\put(4,0){doubly-}
\put(1,-3){displaced-L}
\end{picture}}
\raisebox{0mm}{\setlength{\unitlength}{1mm}
\thicklines
\begin{picture}(20,12)
\put(8,11){\circle{2}}
\put(14,11){\circle*{2.5}}
\put(8,4){\line(1,0){6}}
\put(14,4){\line(0,1){6}}
\put(8,4){\line(0,1){6}}
\put(4,0){triply-}
\put(2,-3){displaced-U}
\end{picture}}
\raisebox{0mm}{\setlength{\unitlength}{1mm}
\thicklines
\begin{picture}(20,15)
\put(7,5){\circle{2}}
\put(11,7){\circle*{2.5}}
\put(7,12){\line(1,0){8}}
\put(7,6){\line(0,1){6}}
\put(15,12){\line(-3,-4){3.0}}
\put(4,0){triply-}
\put(2,-3){displaced-O}
\end{picture}} }
\caption[captab]{The spatial arrangements of the quark-antiquark meson operators.
In the illustrations, the smeared quarks fields are depicted by solid circles, 
each hollow circle indicates a smeared ``barred'' antiquark field, and the 
solid line segments indicate covariant displacements.
\label{fig:mesops}}
\end{figure}

Thus, these authors advocate a three-stage approach to constructing
hadron operators.  First, basic building blocks are
chosen.  These are taken to be smeared covariantly-displaced quark fields
\begin{equation}
\bigl(\widetilde{D}^{(p)}_j\ \widetilde{\psi}\bigr)^A_{a\alpha},
 \  \bigl(\widetilde{\overline{\psi}}  
   \ \widetilde{D}^{(p)\dagger}_j \bigr)^A_{a\alpha},
 \qquad -3\leq j\leq 3,
\label{eq:blocks}
\end{equation}
where $A$ is a flavor index, $a$ is a color index, $\alpha$ is a
Dirac spin index, and the $p$-link gauge-covariant displacement
operator in the $j$-th direction is defined by
\begin{equation}
 \widetilde{D}_j^{(p)}(x,x^\prime) =
 \widetilde{U}_j(x)\ \widetilde{U}_j(x\!+\!\hat{j})\dots 
   \widetilde{U}_j(x\!+\!(p\!-\!1)\hat{j})\delta_{x^\prime,x+p\hat{j}},
\qquad \widetilde{D}_0^{(p)}(x,x^\prime) = \delta_{xx^\prime},
\end{equation}
for $j=\pm 1,\pm 2,\pm 3$ and $p\geq 1$, and where $j=0$ defines
a zero-displacement operator to indicate no displacement. 
Next, {\em elemental} operators $B^{F}_i(t,\bm{x})$ are devised
having the appropriate flavor structure characterized by isospin, strangeness,
{\it etc.}, and color structure constrained by gauge invariance.
For zero momentum states, translational invariance is imposed:
$B^{F}_i(t)=\sum_{{\bm{x}}}  B^{F}_i(t,\xvec).$
Finally, group-theoretical projections are applied to
obtain operators which transform irreducibly under 
all lattice rotation and reflection symmetries:
\begin{equation}
  {\cal B}_{i}^{\Lambda\lambda F}(t)
 = \frac{d_\Lambda}{g_{O_h^D}}\sum_{R\in O_h^D} 
  \Gamma^{(\Lambda)}_{\lambda\lambda}(R)\ U_R\ B^F_i(t)\ U_R^\dagger,
\label{eq:project}
\end{equation}
where $O_h^D$ is the double group of $O_h$, $R$ denotes an element of $O_h^D$,
$g_{O_h^D}$ is the number of elements in $O_h^D$, and $d_\Lambda$ is the
dimension of the $\Lambda$ irreducible representation.  Projections onto
both the single-valued and double-valued irreps of $O_h$ require using the 
double group $O_h^D$ in Eq.~(\ref{eq:project}). 
Given $M_B$ elemental $B^F_i$ operators, many of the projections in 
Eq.~(\ref{eq:project}) vanish or lead to linearly-dependent operators,
so one must then choose suitable linear combinations of the projected operators
to obtain a final set of independent baryon operators.
Thus, in each symmetry channel, one ends up with a set of $r$ operators 
given in terms of a linear superposition of the $M_B$ elemental operators.
The different spatial configurations (see Fig.~\ref{fig:operators} for
the baryon configurations and Fig.~\ref{fig:mesops} for the meson
configurations) yield operators which effectively
build up the necessary orbital and radial structures of the hadron
excitations.  The design of these operators is such that a large number
of them can be evaluated very efficiently, and components in their
construction can be used for both meson, baryon, and multi-hadron
computations.

\begin{figure}[t]
\begin{center}
\includegraphics[width=4.0in, bb=0 40 567 559]{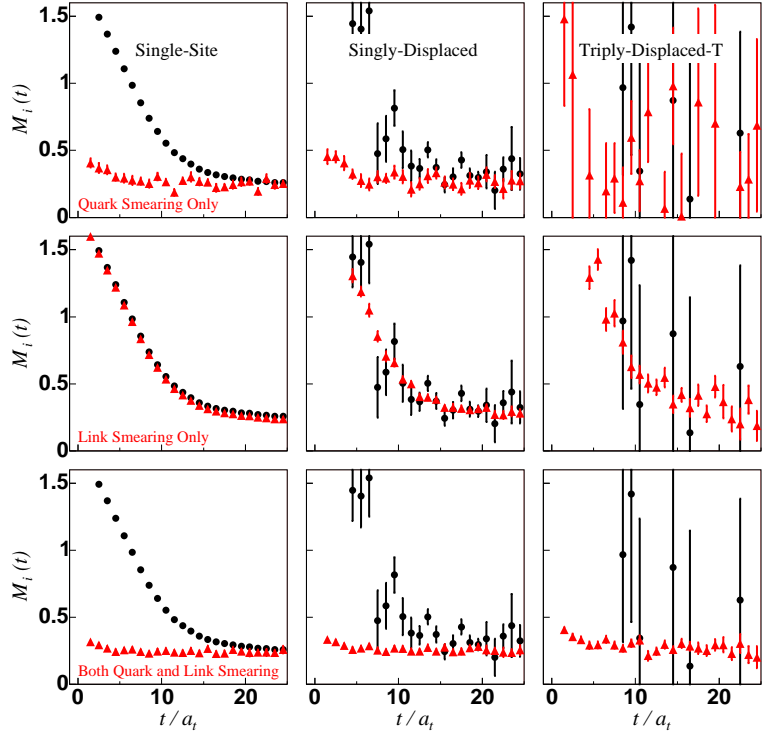}
\end{center}
\caption{Effective masses $M(t)$ for unsmeared (black circles) and smeared 
(red triangles) operators $O_{SS},\ O_{SD},\ O_{TDT}$, which are
representative single-site, singly-displaced, and triply-displaced-T 
nucleon operators, respectively.
Top row: only quark-field smearing $n_\sigma=32,\ \sigma_s=4.0$ is used. Middle row: 
only link-variable smearing $n_\rho=16,\ n_\rho\rho=2.5$ is applied.  
Bottom row: both quark and link smearing $n_\sigma=32,\ \sigma_s=4.0, 
\ n_\rho=16,\ n_\rho\rho=2.5$ are used, dramatically improving the signal for all
three operators. Results are based on 50 quenched configurations on a 
$12^3\times 48$ anisotropic lattice using the Wilson action with $a_s \sim 0.1$ fm,
$a_s/a_t \sim 3.0$.\label{fig:smearing}}
\end{figure}

Finding appropriate smearing parameters is a first crucial part 
of any hadron spectrum calculation.  Fig.~\ref{fig:smearing} demonstrates 
that {\em both} quark-field and link-field smearing are needed in order for
spatially-extended baryon operators to be useful\cite{lat2005}.
It is important to use the smeared links when smearing the quark field.
Link smearing dramatically reduces the statistical errors in the
correlators of the displaced operators, while quark-field smearing
dramatically reduces the excited-state contamination.

\begin{figure}[t]
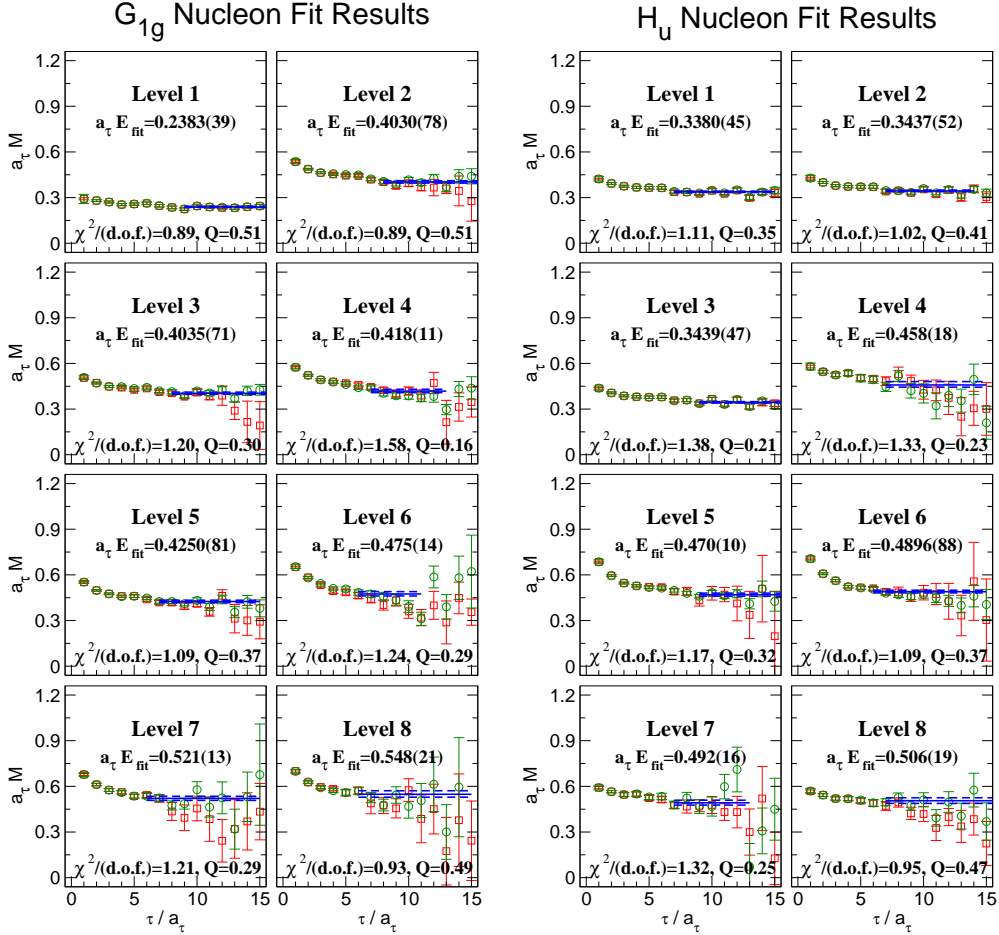

\begin{center}
\includegraphics[width=2.5in,bb=7 40 332 636]{fit-G1g}\hspace{4mm}
\includegraphics[width=2.5in,bb=7 40 332 633]{fit-Hu}
\end{center}
\caption{Effective masses for the lowest eight levels in the $G_{1g}$
channel (left) and $H_u$ channel (right).  The green circles are
fixed-coefficient effective masses, whereas the red squares are
the principal effective masses.  Fit values are shown by the blue lines.
Results are based on 200 quenched configurations on a 
$12^3\times 48$ anisotropic lattice using the Wilson gauge and quark
actions with $a_s \sim 0.1$ fm,
$a_s/a_t \sim 3.0$ and $m_\pi\sim 700$~MeV.
\label{fig:effmasses}}
\end{figure}

The above approach to designing hadron and multi-hadron interpolating
fields leads to a very large number of operators.  It is not
feasible to do spectrum computations using all of the operators 
so designed; for example, in the $G_{1g}$ symmetry channel for nucleons,
the above procedure leads to 179 operators.  It is necessary to \textit{prune}
down the number of operators.  Six months of exploratory testing and trials
led to the following guideline: noise is the enemy, so a procedure that
keeps a variety of operators while minimizing the effects of noise
works best.  Some operators are intrinsically noisy and must be removed.
In addition, a set of operators, each with little intrinsic noise, can allow
noise to creep in if they are not sufficiently independent of one another.

In Ref.~\cite{adamthesis}, the following procedure is advocated. (1) First, 
remove operators with excessive intrinsic noise.  This can be done by
examining the diagonal elements of the correlation matrix and discarding
those operators whose self-correlators have relative errors above some
threshold for a range of temporal separations.  Of course, this requires
a low-statistics Monte Carlo computation on a reasonably small lattice.
(2) Second, prune within operator types (single-site, singly-displaced, 
\textit{etc.}) based on the condition number of the submatrices
\[
   \widehat{C}_{ij}(t) = \frac{C_{ij}(t)}{\sqrt{C_{ii}(t)C_{jj}(t)}},
\qquad t=a_t.
\]
The condition number is taken to be the ratio of the largest eigenvalue
over the smallest eigenvalue.  A value near unity is ideal.
For each operator type, the set of about six operators which yields
the lowest condition number of the above submatrix is retained.
(3) Lastly, prune across all operator types based again on the condition
number of the remaining submatrix as defined above.  In this last step,
the goal is to choose about 16 operators, keeping two or three of each type,
such that a condition number reasonably close to unity is obtained.  As long
as a good variety of operators is retained, the resulting spectrum seems
to be fairly independent of the exact choice of operators at this stage.
Eigenvectors from a variational study of the operators can also be
used to fine tune the choice of operators.

Calculations in Ref.~\cite{adamthesis} using about 16 operators in all irreps 
for the nucleon channel on only 200 configurations were very successful.  The
extraction of 8 energy levels in each irrep was possible, which was a major 
milestone achieved.  A determination of the hadron spectrum requires the ability 
to extract several excited energy levels, and up until that time, it was
not known whether or not extracting more than one or two levels would be possible.
Fig.~\ref{fig:effmasses} shows the signal quality in the $G_{1g}$ and $H_u$ 
irreps for the nucleon excitations from that first calculation.  Similar
calculations for the $\Delta$ resonance spectra have also been 
achieved\cite{peru1,peru2}.  Comparison of these results with experiment is not
justified since the quenched approximation was used, an unphysically
large $u,d$ quark mass was used, and the lattice volume is too small.

It is my strong opinion that the use of correlation matrices is the
best way to extract excited-state energies reliably.  However, there are
efforts to deduce information about excited states from single
correlation functions.  Bayesian statistics have been used\cite{bayesian1,
bayesian2}, as well as maximum entropy methods\cite{mem1,mem2}.
A novel evolutionary fitting method has been proposed\cite{evolution},
and a new method based on statistical concepts which relies heavily on
simulation techniques was presented at this conference\cite{dina}.

\section{Stochastic estimates of many-to-many quark propagators with source
 dilution variance reduction}

To study a particular eigenstate of interest, all eigenstates lying below that
state must first be extracted, and as the pion gets lighter in lattice QCD 
simulations, more and more multi-hadron states will lie below the excited
resonances.  Consider a baryon at rest.  An appropriate quantum operator
for a baryon at rest typically has the form
\begin{equation}
  B(\pvec=0,t) = \frac{1}{V}\sum_{\rvec} \varphi_B(\rvec,t),
\end{equation}
where $V$ is the volume of the lattice and $\varphi_B(\rvec,t)$ is an
appropriate localized interpolating field.  In the above equation, the summation
over spatial lattices makes the operator translationally invariant, producing
a zero momentum state.  A baryon correlator, thus, has a double summation
over spatial sites:
\begin{equation}
 \langle 0\vert B(\pvec=0,t)\overline{B}(\pvec=0,0)\vert 0\rangle
 = \frac{1}{V^2}\sum_{\xvec,\yvec}\langle 0\vert \varphi_B(\xvec,t)
\overline{\varphi}_B(\yvec,0)\vert 0\rangle.
\end{equation}
Evaluating the above correlator requires computing the $\yvec\rightarrow\xvec$
element of the quark propagators.  In other words, the quark propagators from
all spatial sites $\yvec$ on time slice $t=0$ to all spatial sites $\xvec$
on later time slice $t>0$ must be known.  Computing all such elements
of the propagators exactly is not possible (except on very small lattices).
In the example above, this problem can be circumvented by appealing to translational
invariance to limit the summation over the source site to a single site:
\begin{equation}
 \langle 0\vert B(\pvec=0,t)\overline{B}(\pvec=0,0)\vert 0\rangle
 = \frac{1}{V}\sum_{\xvec}\langle 0\vert \varphi_B(\xvec,t)
\overline{\varphi}_B(\bm{0},0)\vert 0\rangle.
\end{equation}
However, a \textit{good} baryon-meson operator of total zero momentum
typically has the form
\begin{equation}
 B(\pvec,t)M(-\pvec,t)=\frac{1}{V^2}\sum_{\xvec,\yvec}\varphi_B(\xvec,t)
\varphi_M(\yvec,t)e^{i\pvec\cdot(\xvec-\yvec)},
\end{equation}
where $\varphi_M(\yvec,t)$ is a localized interpolating field for a meson.
In the evaluation of the temporal correlations of such a multi-hadron
operator, it is not possible to completely remove all summations over the
source site.  Hence, the need for estimates of the quark propagators from 
all spatial sites on a time slice to all spatial sites on another time slice
cannot be sidestepped.  Ultimately, some correlators will involve
disconnected diagrams which necessarily involve all-to-all quark propagators.
Hence, all-to-all (or many-to-many) quark propagators are becoming mandatory,
and some way of stochastically estimating them is needed.

Random noise vectors $\eta$ whose expectations satisfy
$E(\eta_i)=0$ and $E(\eta_i\eta_j^\ast)=\delta_{ij}$ are useful for 
stochastically estimating the inverse of a large matrix $M$ 
as follows.  Assume that for each of $N_R$ 
noise vectors, we can solve the following
linear system of equations: $M X^{(r)}=\eta^{(r)}$ for $X^{(r)}$.
Then $X^{(r)}=M^{-1}\eta^{(r)}$, and
\begin{equation}
   E( X_i \eta_j^\ast ) = E( \sum_k M^{-1}_{ik}\eta_k \eta_j^\ast )
  = \sum_k M^{-1}_{ik}
  E(\eta_k \eta_j^\ast) = \sum_k M^{-1}_{ik} \delta_{kj} = M^{-1}_{ij}.
\end{equation}
The expectation value on the left-hand can be approximated using the
Monte Carlo method.  Hence, a Monte Carlo estimate of $M_{ij}^{-1}$
is given by
\begin{equation}
  M_{ij}^{-1} \approx \lim_{N_R\rightarrow\infty}\frac{1}{N_R}
 \sum_{r=1}^{N_R} X_i^{(r)}\eta_j^{(r)\ast}, \qquad
\mbox{where $MX^{(r)}=\eta^{(r)}$.}
\end{equation}
Unfortunately, this equation usually produces stochastic 
estimates with variances which are much too large to be useful.

Progress is only possible if stochastic estimates of the quark
propagators with reduced variances can be made.  Techniques of
\textit{diluting} the noise vectors have been developed which
accomplish such a variance 
reduction\cite{dilute1,dilute2,dilute3,dilute4,dilute5,dilute6}.
A given dilution scheme can be viewed as the application of a complete
set of projection operators.  To see how dilution works, consider a general 
$N\times N$ matrix $M$ having matrix elements $M_{ij}$.  
Define some complete set of $N\times N$ projection matrices $P^{(a)}$ which 
satisfy
\begin{equation}
 P^{(a)}P^{(b)}=\delta^{ab}P^{(a)},\qquad \sum_a P^{(a)}=1,
 \qquad P^{(a)\dagger}=P^{(a)}.
\label{eq:projectors}
\end{equation}
Then observe that
\begin{eqnarray}
  M_{ij}^{-1}&=&M_{ik}^{-1}\delta_{kj}=\sum_a M_{ik}^{-1}P^{(a)}_{kj}
  =\sum_a M_{ik}^{-1}P^{(a)}_{kk^\prime}P^{(a)}_{k^\prime j}
  =\sum_a M_{ik}^{-1}P^{(a)}_{kk^\prime}\delta_{k^\prime j^\prime} P^{(a)}_{j^\prime j}
 \nonumber \\
  &=&\sum_a M_{ik}^{-1}P^{(a)}_{kk^\prime}E(\eta_{k^\prime}\eta^\ast_{j^\prime}) 
    P^{(a)}_{j^\prime j}
   =\sum_a M_{ik}^{-1}E\left(P^{(a)}_{kk^\prime}\eta_{k^\prime}
    \eta^\ast_{j^\prime}P^{(a)}_{j^\prime j}\right). 
\end{eqnarray}
Define
\begin{equation}
  \eta^{[a]}_k=P^{(a)}_{kk^\prime}\eta_{k^\prime} ,\qquad
  \eta^{[a]\ast}_j=\eta^\ast_{j^\prime}P^{(a)}_{j^\prime j}
    =P^{(a)\ast}_{j j^\prime}\eta^\ast_{j^\prime}
\end{equation}
and further define $X^{[a]}$ as the solution of
\begin{equation}
   M_{ik}X^{[a]}_k=\eta^{[a]}_i,
\end{equation}
then we have
\begin{equation}
   M_{ij}^{-1}=\sum_a M_{ik}^{-1} E(\eta^{[a]}_k \eta^{[a]\ast}_j)
      =\sum_a E(X^{[a]}_i\eta^{[a]\ast}_j).
\label{eq:diluted}
\end{equation}
Although the expected value of $\sum_a \eta^{[a]}_k
\eta^{[a]\ast}_j$ is the same as $\eta_k\eta^\ast_j$, the \textit{variance} 
of $\sum_a \eta^{[a]}_k\eta^{[a]\ast}_j$ is significantly smaller than that 
of $\eta_k\eta^\ast_j$.  For both $Z_4$ and $U(1)$ noise, we
have
\[  Var({\rm Re}(\eta_i\eta_j^\ast))=Var({\rm Im}(\eta_i\eta_j^\ast))=
 \textstyle\frac{1}{2}(1-\delta_{ij}).\]
Although the variance is zero for $i=j$, there is a significant variance 
for all $i\neq j$.  The dilution projections ensure \textit{exact zeros} 
for many of the off-diagonal elements, instead of values that are only
statistically zero.  In other words, many of the $i\neq j$ elements
become exactly zero.

Of course, the effectiveness of the variance reduction depends on the
projectors chosen.  A particularly important dilution scheme for measuring 
temporal correlations in hadronic quantities is ``time dilution'' where the noise
vector is broken up into pieces which only have support on a single time
slice:
\begin{equation}
  P_{a\alpha;b\beta}^{(B)}(\xvec,t;\yvec,t^\prime)
 =\delta_{ab}\delta_{\alpha\beta}\delta_{\xvec\yvec}
 \delta_{Bt}\delta_{Bt^\prime},
\qquad B=0,1,\cdots, N_t-1, \quad\mbox{(time dilution)},
\end{equation}
where $N_t$ is the number of time slices on the lattice, $a,b$ are
color indices, and $\alpha,\beta$ are spin indices.  Spin and color
dilution are two other easy-to-implement schemes:
\begin{eqnarray}
  P_{a\alpha;b\beta}^{(B)}(\xvec,t;\yvec,t^\prime)
 &=&\delta_{ab}\delta_{B\alpha}\delta_{B\beta}\delta_{\xvec\yvec}\delta_{tt^\prime},
\qquad B=0,1,2,3, \quad\mbox{(spin dilution)},\\
  P_{a\alpha;b\beta}^{(B)}(\xvec,t;\yvec,t^\prime)
 &=&\delta_{Ba}\delta_{Bb}\delta_{\alpha\beta}\delta_{\xvec\yvec}\delta_{tt^\prime},
\qquad B=0,1,2, \quad\mbox{(color dilution)}.
\end{eqnarray}
Various spatial dilution schemes are possible, too.  For example, even-odd
dilutions are simple to implement.  The above dilution projectors can also be
combined to make hybrid schemes.

\begin{figure}
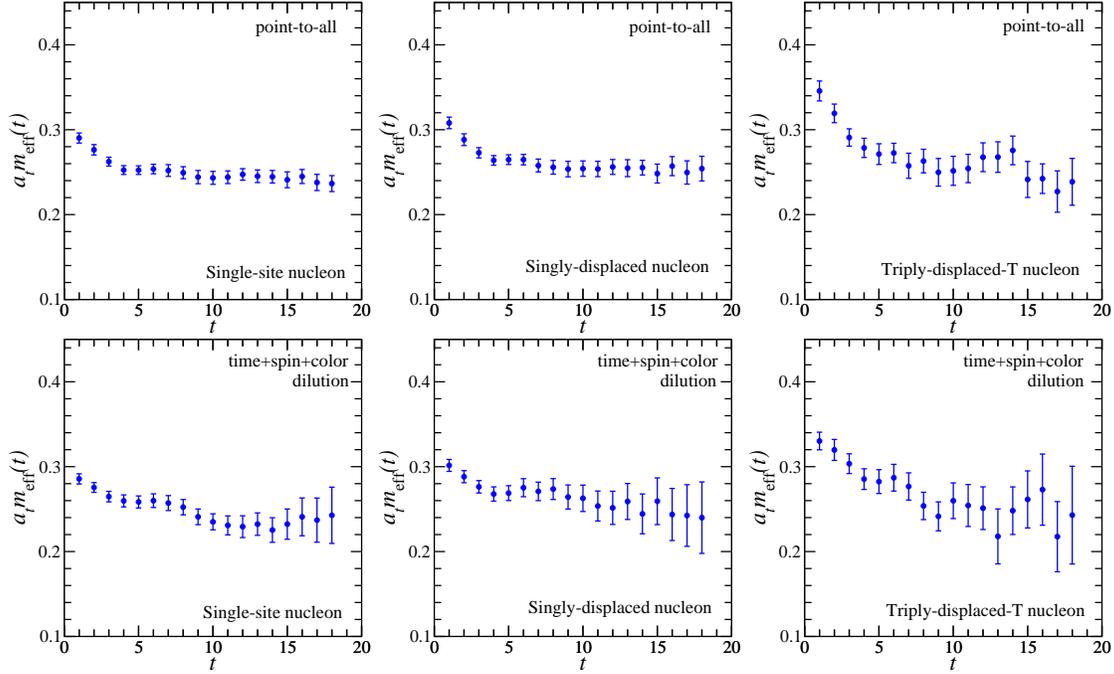

\begin{center}
\includegraphics[width=1.9in,bb=11 41 534 523]{SS_pt2all_effmass.eps}
\includegraphics[width=1.9in,bb=11 41 534 523]{SD_pt2all_effmass.eps}
\includegraphics[width=1.9in,bb=11 41 534 523]{TDT_pt2all_effmass.eps}\\
\includegraphics[width=1.9in,bb=11 41 534 523]{SS_time+spin+color_effmass.eps}
\includegraphics[width=1.9in,bb=11 41 534 523]{SD_time+spin+color_effmass.eps}
\includegraphics[width=1.9in,bb=11 41 534 523]{TDT_time+spin+color_effmass.eps}
\end{center}
\caption[dilute1]{
(Upper row) Effective masses for a single-site (left), singly-displaced (middle),
and triply-displaced-T (right) nucleon operator using quark propagators evaluated
with the standard point-to-all method. (Lower row) The effective masses
for the same nucleon operators but using stochastic quark propagators with
time+spin+color dilution.  Without dilutions, the errors in these effective
masses would be orders of magnitude larger.  An effective mass defined
using a time separation $3a_t$ is used in these plots.  These results used
100 quenched configurations on an anisotropic $12^3\times 48$ lattice with a
Wilson fermion and gauge action.
\label{fig:Deffmasses}}
\end{figure}

Before presenting tests of these different dilution schemes, an
important remark about the use of stochastic quark propagators should be
mentioned.  The use of Eq.~(\ref{eq:diluted}) to approximate quark propagators
leads to a very desirable source-sink factorization.  Consider a baryon
correlator of the form
\begin{equation}
 C_{l\overline{l}} = c_{ijk}^{(l)} c^{(\overline{l})\ast}_{\overline{i}\overline{j}
\overline{k}} Q^{(A)}_{i\overline{i}}Q^{(B)}_{j\overline{j}}
Q^{(C)}_{k\overline{k}},
\end{equation}
where $Q^{(A)}$ denotes a quark propagator of flavor $A$ and 
all other quark indices have been combined into a single index $i$ or 
$j$, and so on.  Stochastic estimates of this correlator using
Eq.~(\ref{eq:diluted}) lead to the form
\begin{equation}
 C_{l\overline{l}} = \frac{1}{N_R}\sum_r \sum_{d_Ad_Bd_C}
 c_{ijk}^{(l)} c^{(\overline{l})\ast}_{\overline{i}\overline{j}
\overline{k}} 
\left(\varphi_i^{(Ar)[d_A]}\eta_{\overline{i}}^{(Ar)[d_A]\ast}\right)
\left(\varphi_j^{(Br)[d_B]}\eta_{\overline{j}}^{(Br)[d_B]\ast}\right)
\left(\varphi_k^{(Cr)[d_C]}\eta_{\overline{k}}^{(Cr)[d_C]\ast}\right),
\end{equation}
where $r$ labels the noise vectors, $d_A,d_B,d_C$ are the dilution
indices, $\eta$ are the noise vectors, and $\varphi$ are the solution
vectors.  If one defines
\begin{eqnarray}
 \Gamma_l^{(r)[d_Ad_Bd_C]} &=& c_{ijk}^{(l)} 
 \varphi_i^{(Ar)[d_A]} \varphi_j^{(Br)[d_B]} \varphi_k^{(Cr)[d_C]},
\label{eq:source}\\
 \Omega_l^{(r)[d_Ad_Bd_C]} &=& c_{ijk}^{(l)} 
 \eta_{i}^{(Ar)[d_A]}\eta_{j}^{(Br)[d_B]}\eta_{k}^{(Cr)[d_C]},
\label{eq:sink}
\end{eqnarray}
then the baryon correlator becomes a glorified dot product of the source
vector with the sink vector:
\begin{equation}
 C_{l\overline{l}} = \frac{1}{N_R}\sum_r \sum_{d_Ad_Bd_C}
 \Gamma_l^{(r)[d_Ad_Bd_C]}\Omega_{\overline{l}}^{(r)[d_Ad_Bd_C]\ast}.
\end{equation}
The source and sink vectors in Eqs.~(\ref{eq:source}) and (\ref{eq:sink})
can be separately evaluated for a variety of operators, and the dot
product applied afterwards to evaluate the matrix of correlation
functions.  Different $ABC$ permutations of the noise vectors must be 
stored in order to accommodate all needed Wick contractions.  The
use of stochastic all-to-all quark propagators has led to an enormous
simplification of the effort required to compute the hadron correlation
matrices through this source-sink factorization.  Another advantage of this
approach is the fact that, given suitable non-zero momenta, these same 
baryon and meson operators can be combined later to make multi-hadron
operators.

The effectiveness of stochastically-estimated all-to-all quark propagators using
diluted noise vectors is demonstrated in Fig.~\ref{fig:Deffmasses}.
This figure compares the effective masses for a single-site,
singly-displaced, and triply-displaced-T nucleon operator using
quark propagators evaluated with the conventional point-to-all method (top row)
and with the all-to-all stochastic method including time+spin+color dilutions (bottom
row).  The fact that these effective masses have comparable errors
indicates that the stochastic method with suitable dilutions has not
introduced any appreciable noise into the final mass extractions.

\begin{figure}
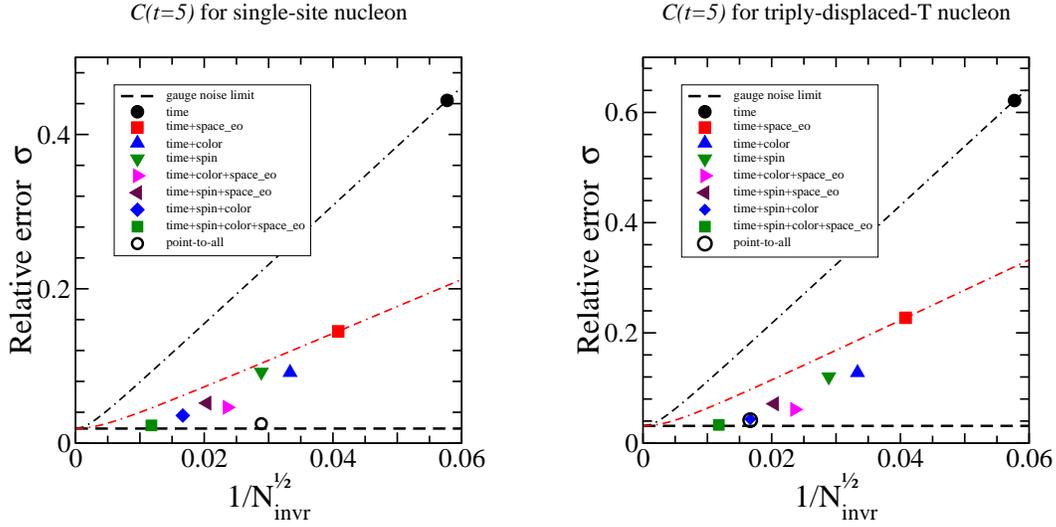

\begin{center}
\includegraphics[width=2.5in,bb=17 5 547 581]{SS_C5_dilutions.eps}
\hspace*{10mm}
\includegraphics[width=2.5in,bb=17 5 547 581]{TDT_C5_dilutions.eps}
\end{center}
\caption[dilute]{
(Left) The relative errors in the correlation function of a single-site 
nucleon operator for temporal separation $t=5a_t$ evaluated using
stochastically-estimated quark propagators with different dilution
schemes against $1/N_{\rm inv}^{1/2}$, where $N_{\rm inv}$ is the
number of Dirac matrix inversions required. The open circle shows the
point-to-all error, and the horizontal dashed line shows the gauge-noise
limit.  The black (red) dashed-dotted line shows the decrease in error
expected by simply increasing the number of noise vectors, starting 
from the time (time + even/odd-space) dilution point.  (Right)
Same as the left plot, except for a triply-displaced-T nucleon operator.
These results used 100 quenched configurations on an anisotropic
$12^3\times 48$ lattice with a Wilson fermion and gauge action.
\label{fig:dilutions}}
\end{figure}

A comparison of different dilution schemes has been presented at this
conference\cite{bulava}.  Fig.~\ref{fig:dilutions} shows the relative errors
in the correlation functions of a single-site and a triply-displaced-T
nucleon operator for temporal
separation $t=5a_t$ evaluated using stochastically-estimated quark propagators
with different dilution schemes against $1/N_{\rm inv}^{1/2}$, where 
$N_{\rm inv}$ is the number of matrix inversions required. These results were
obtained using 100 quenched configurations on an anisotropic $12^3\times 48$
lattice with a Wilson fermion and gauge action.  The open circles show the
point-to-all errors, and the horizontal dashed lines show the gauge-noise
limits.  The black (red) dashed-dotted lines show the decrease in error
expected by simply increasing the number of noise vectors, starting 
from the time (time + even/odd-space) dilution points.  These computations
are dominated by the inversions of the Dirac matrix, so using the number of
matrix inversions $N_{\rm inv}$ to compare computational efforts is 
reasonably fair.  The advantage in using increased dilutions compared to an
increased number of noise vectors with only time dilution is evident in
the plots.  However, this advantage quickly diminishes after
time + even/odd-space dilution, or time+color, or time+spin dilution.
Note that time+spin+color+even/odd-space dilution yields an error
comparable with the gauge-noise limit using only a single noise vector!
First results for multi-hadron operators were also presented at
this conference\cite{juge}.

These encouraging results demonstrate that the inclusion of good multi-hadron
operators will certainly be possible using stochastic all-to-all
quark propagators with diluted-source variance reduction.  In fact,
just before this conference, the authors of Ref.~\cite{nucleons2} began exploring
a new method that might allow nearly-exact determinations of many-to-many
quark propagators without introducing any noise vectors at all.  The method
exploits a novel, cleverly-devised choice of quark-field smearing to facilitate
the nearly-exact computations.  Details and tests of this method should appear
very soon.

\section{Summary and outlook}

This talk discussed the key issues and challenges in exploring excited
hadrons in lattice QCD. The importance of multi-hadron operators
and the need for all-to-all quark propagators were emphasized.  The
challenge of dealing with unstable states (resonances) in a
box was outlined, and the technology associated with extracting excited
stationary-state energies, including operator design and field smearing,
was detailed.  Efforts in variance reduction of stochastically-estimated
all-to-all quark propagators using source dilutions were described, and 
results on excited hadrons which appeared during the last year were 
summarized.

Given the major experimental efforts to map out the QCD resonance spectrum, 
such as Hall B and the proposed Hall D at Jefferson Lab, ELSA associated 
with the University of Bonn, COMPASS at CERN, PANDA at GSI, and BESIII 
in Beijing, there is a great need for \textit{ab initio} determinations of
such states in lattice QCD. The exploration of excited hadrons in lattice
QCD is well underway.

This work was supported by the National Science Foundation through awards
PHY 0653315 and PHY 0510020.

\end{document}